# Acoustic-induced strong interaction between two periodically patterned elastic plates

Chunyin Qiu, Shengjun Xu, Manzhu Ke, and Zhengyou Liu*

Key Laboratory of Artificial Micro- and Nano-structures of Ministry of Education and School of Physics and Technology, Wuhan University, Wuhan 430072, China

**Abstract:**

We study the acoustic-induced interactions between a pair of identical elastic plates perforated with periodical structures. Tremendous mutual forces, both repulsions and attractions, have been observed in subwavelength regime. The dramatic effect stems from the resonant enhancement of the local field sandwiched between the double plates. The parameter-sensitivity of the magnitude and the sign of the interaction (i.e. repulsion or attraction) depend directly on the vibration morphology of the resonant mode. In practical applications, the sign of the interaction can be switched by controlling the external frequency. Both the adjustable magnitude and the switchable sign of the contactless interaction endow this simple and compact double-plate structure with great potential in ultrasonic applications.

**PACS numbers:** 43.25.Qp, 43.20.El, 43.20.Ks

*Author to whom correspondence should be addressed. Email: zyliu@whu.edu.cn

**I. Introduction**

Acoustic waves can exert forces on illuminated objects via momentum exchanges between the sound field and objects [1-3]. Comparing with other contactless manipulations, the acoustic radiation forces (ARFs) necessitate much less power and exhibit less destruction to objects. Besides, the ARFs work for nearly any type of objects regardless to the optic and electronic properties. These merits enable the ARFs to be highly competent in a wide range of applications [4], such as in particle trapping and patterning, and acoustic levitations [4-11]. The properties of ARFs are closely related to the characteristics of acoustic sources and objects simultaneously. Via introducing exotic acoustic sources, recently, many fascinating phenomena on ARFs have been demonstrated [12-18], where prominent representatives can be referred to the acoustic-induced rotating [12-15] and pulling effects [16-19]. Note that a majority of the existed ARF studies focused on objects consisting of monomers or very dilute suspensions associated with negligible particle-particle interactions. In some occasions, however, the objects are essentially multibody systems and the acoustic-induced interaction among the monomers turns out to be innegligible. A remarkable case is the acoustic-induced self-organization effect in a cluster of air bubbles immersed in liquids [20,21].

In recent decades, artificial structures for optic [22] and acoustic [23] waves have been attracting a significant interest because of numerous unprecedented wave responses. These exotic wave responses, especially the near-field couplings, could bring a rich diversity of interactions induced by optic waves or acoustic waves. In optics, the optically induced interactions in various coupled artificial structures are receiving a fast growing attention, due to the great potentials in sensing, switching and other optical devices [24-34]. The interactions can be flexibly tuned to be either repulsive or attractive, and also can be tuned much larger than that acting on each monomer individually or acting on the whole system through near-field resonant couplings between monomers. Similar phenomena and applications in coupled artificial structures can be anticipated in acoustics as well. A recent study states that airborne ARFs exerting on a rigid wall can be remarkably amplified through the aid of the resonant coupling with an adjunctive metamaterial slab [35]. The enhanced pushing force can even overcome the gravity and lift a macroscopic object, which constitutes a striking contrast to the previous acoustic manipulations on micro-sized particles. It is worth pointing out that only the repulsive interaction has been realized in that specific airborne sound system, where the coupled solid structure is considered as rigid with respect to air ambience.

As to be shown below, both the sign (i.e. attraction or repulsion) and the magnitude of the acoustic-induced mutual force (AIMF) is determined by the morphology of the resonant acoustic field. Recent studies on the extraordinary acoustic transmissions through plate-like structures state that various intrinsic modes in the solid plates can be resonantly excited through the scattering of periodical structures, once the solid's elasticity is fully taken into account [36,37]. These resonances correspond to a variety of acoustic field morphologies. It is believed that if considering coupled plate structures further, more abundant resonant morphologies can be achieved to produce numerous interesting interaction behaviors. Here we consider a pair of identical elastic plates patterned with one-dimensional (1D) periodic array of solid bumps. Besides the enhanced repulsive interaction that was realized previously, our results demonstrate extremely strong attractive interaction as well. The latter is produced by the resonant excitation of the anti-phase-coupled flexural modes, in which the tangential velocity localized in between the plates is drastically amplified. Prospective applications of such enhanced and adjustable contactless interactions can be anticipated through such simple and compact plate structures, e.g. in designing acoustic sensors and switches.

The remainder of this paper is organized as follows. In Sec. II. we present the theoretical framework to evaluate the ARF and AIMF for the structured double-plate system. In Sec. III. we present the numerical results for the transmission and AIMF spectra, where the latter demonstrates greatly enhanced repulsive and attractive interactions at different resonant wavelengths. It is followed by an explanation presented in Sec. IV., based on a comprehensive analysis of the vibration morphologies at resonances. The parameter-sensitivities of these enhanced interactions are explored in Sec. V. Finally, a brief is made in Sec. VI.

## II. Theoretical Model for Evaluating ARFs and AIMFs

As illustrated in Fig. 1, the system under consideration is a pair of identical steel plates (of thickness $h$), each patterned with a 1D periodical array (at a pitch $L$) of square bumps (with side-length $a$) on a single side. The structured plates are immersed in water and placed in parallel at a distance $d$. Without losing generality, here we take the geometry parameters $a=d=h=L/10$. The material parameters used are listed below: mass density $\rho_s = 7.67 g/cm^3$, longitudinal velocity $c_l = 6.01 km/s$ and transverse velocity $c_t = 3.23 km/s$ for steel; mass density $\rho_0 = 1.0 g/cm^3$ and sound velocity $c_0 = 1.49 km/s$ for water. In this work, we focus on the external incidence of a plane wave (with pressure amplitude $p_0$) impinging normally onto the sample (along +y direction). The excited acoustic field described by the first-order pressure $p$ and velocity $\mathbf{u}$ can be numerically simulated by the finite-element solver (COMSOL MULTIPHYSICS), which is used to calculate the transmission coefficient and the ARF depicted below.

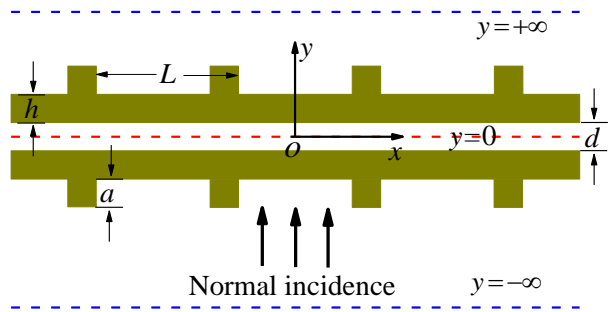

FIG. 1. (color online). Schematic diagram for a water-immersed parallel-plate system, which consists of two identical steel plates patterned with 1D periodic array of square bumps. The three dash lines denote the integral surfaces for calculating the ARFs exerting on both plates.

For an ideal fluidic background, i.e., ignoring its viscosity and absorption, the

ARF exerting on an object can be evaluated by an arbitrary integral contour enclosing the object, i.e.,

$$\mathbf{F} = -\oint \langle \mathbf{\Pi} \rangle \cdot \mathrm{d}\mathbf{S}, \tag{1}$$

where $\langle \mathbf{\Pi} \rangle = \left[ |p|^2 / (4\rho_0 c_0^2) - \rho_0 |\mathbf{u}|^2 / 4 \right] \mathbf{I} + \rho_0 \mathbf{u}^* \mathbf{u} / 2$ is the time-averaged Brillouin radiation stress tensor and the differential area $\mathrm{d}\mathbf{S}$ points to its outer normal. As usual here $\mathbf{I}$ stands for a unit tensor and the asterisk denotes the complex conjugation. As shown in Fig. 1 by dashed lines, it is convenient to calculate the ARF exerting on the upper plate by the integral over the middle plane (of the whole structure) $y = 0$ and the far-field boundary $y = +\infty$, and calculate the ARF exerting on the lower plate by $y = 0$ and $y = -\infty$.

At normal incidence it is easy to see that, only the $y$-component of the ARF survives because of the system's symmetry with respect to $x = 0$. Therefore, the vectorial ARF can be described by a scalar quantity, where the positive and negative signs refer to $+y$ and $-y$ directions, respectively. To reduce complexity, in this work we focus on the frequency range below the first-order diffraction, i.e. $\lambda_0 / L > 1$, where $\lambda_0$ is the wavelength in water. In this case the force components contributed from the integrals over the far-field boundaries $y = +\infty$ and $y = -\infty$ can be simply determined by the power transmission coefficient $t$ and reflection coefficient $r$. After scaled by the unit $E_0 S$, the (dimensionless) force components can be evaluated as $Y_t = -t$ and $Y_r = 1 + r$, respectively, where $E_0 = p_0^2 / (2\rho_0 c_0^2)$ is the energy density of the incident plane wave, and $S$ is the area of the corresponding planar plate. Note that without considering absorptions, the total force acting on the whole system can be simply written as $Y_r + Y_t = 2r$, associated with a maximum of 2. It always behaves as a pushing force with respect to the incident plane wave. The conclusions also hold for the case of a single plate.

Now consider the contribution from the integral over the surface $y=0$, whose surface normal points to $-y$ direction. After scaled by $I_0 S$, the dimensionless quantity can be expressed as

$$Y_m = \frac{1}{2L}\int_{-L/2}^{L/2}\left(|p/p_0|^2 + |u_y/u_0|^2 - |u_x/u_0|^2\right)dx. \tag{2}$$

Therefore, the dimensionless ARFs acting on the upper and lower plates can be written as $Y_u = Y_m + Y_t$ and $Y_l = Y_r - Y_m$, respectively, where $u_0 = p_0/(\rho_0 c_0)$ is the velocity amplitude of the incident plane wave, and $u_x$ and $u_y$ correspond to the velocity components along $x$ and $y$ directions, respectively. In some sense, the quantity $Y_m$ can be regarded as the mutual force between the two plates (induced by acoustic waves), where the positive and negative signs correspond to repelling and attracting interactions, respectively. This definition becomes particularly meaningful when the magnitude of $Y_m$ is tuned much larger than the contributions from $Y_t$ and $Y_r$, where ARFs for the upper and lower plates can be approximated as $Y_u \approx Y_m$ and $Y_l \approx -Y_m$. Note that the value of $Y_m$ is indeed independent of the selection of the integral surface. This can be easily derived from the fact that the integral in Eq. (1) contributes a zero ARF if the contour encloses only fluid.

As indicated in Eq. (2), the first two positive terms contribute to repulsion, whereas the third negative term contributes to attraction. For unstructured parallel-plate systems, at normal incidence the transversal velocity $u_x \equiv 0$ and thus the interaction is always repulsive. In Ref. [35], a strong repulsive interaction between the wall and the metamaterial slab has been realized by introducing a resonance to amplify the pressure intensity (i.e. the first term). In this paper, we demonstrate that both the repulsion and attraction can be realized between the double plates by perforating periodical structures (where the wave scatterings from square bumps contribute nonzero transversal velocity now). The interactions especially the attractive interaction can be tuned tremendously large (enhanced by several orders of magnitude), with fully taking advantage of the resonant modes inherent in the elastic plates.

## III. Spectra for the Transmission and Mutual Force

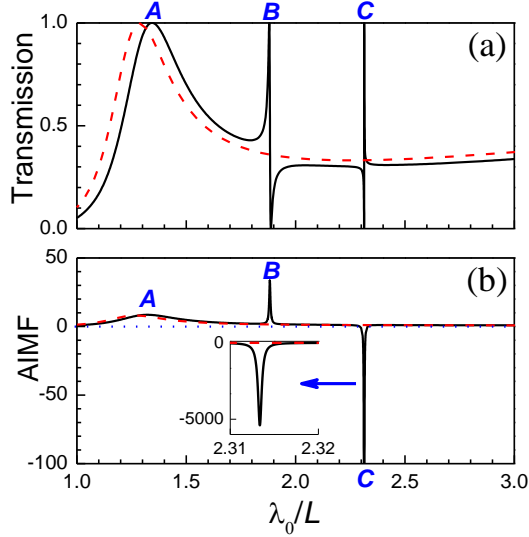

FIG. 2 (color online). Transmission (a) and AIMF (b) spectra for the structured (black solid line) and unstructured (red dashed line) parallel-plate systems. The inset in (b) displays an enlarged view of the spectra near the resonance *C*, which exhibits a huge attractive interaction between the two structured plates.

As shown in Eq. (2), the dimensionless AIMF is usually on the order of one for a moderate acoustic field distribution, i.e. $|p/p_0|^2 \sim 1$, $|u_y/u_0|^2 \sim 1$ and $|u_x/u_0|^2 \sim 1$. To enhance the AIMF it is necessary to attain strong acoustic field localizing in between the parallel plates, which corresponds directly to certain local resonances. Therefore, it is very helpful to study first transmission spectra that can reflect the information of resonances. In Fig. 2(a), we present the power transmission spectrum for the structured parallel-plate system (black solid line), together with that for the unstructured one (red dashed line) for comparison. It is observed from Fig. 2 that for the structured system, there are three striking resonances emerging in the subwavelength region of the transmission spectrum, which are denoted by *A*, *B* and *C*, respectively. The resonance *A* (associated with a broad transmission peak) is less related with the periodical structure since it is manifested in the unstructured system as well, except for a slight deviation in wavelength. The resonances *B* and *C* are unique in the structured system. They are typically fano-type, as featured by the asymmetric line shapes near the narrow peaks and dips.

Now we turn to study the acoustic-induced interaction between the two structured steel plates. In Fig. 2(b) we present the dimensionless AIMF spectrum for the structured parallel-plate system (black solid line), comparing with that for the unstructured one (red dashed line). As predicted, every resonance displayed in the above transmission spectra contributes a sizeable interaction, either repulsive or attractive [38]; away from the resonances, the AIMF recovers to the order of one. Consistent with the band widths of transmission peaks, the AIMF peaks for the resonances *B* and *C* are much sharper than that for the resonance *A*. In particular, the attractive AIMF is extremely strong (as high as ~5000, see inset). Note that the numerical accuracy is safely guaranteed through refining grids in the finite-element simulation.

**IV. Explanations for the Enhanced Repulsive and Attractive Interactions**

To physically understand these remarkable AIMF enhancements, below we study the corresponding acoustic field morphologies at resonances. In Figs. 3(a)-3(c) we plot the normalized near-field distributions for the three resonances *A*-*C*, where the upper panels correspond to pressure fields (normalized by $p_0$) and the lower panels correspond to velocity fields (normalized by $u_0$).

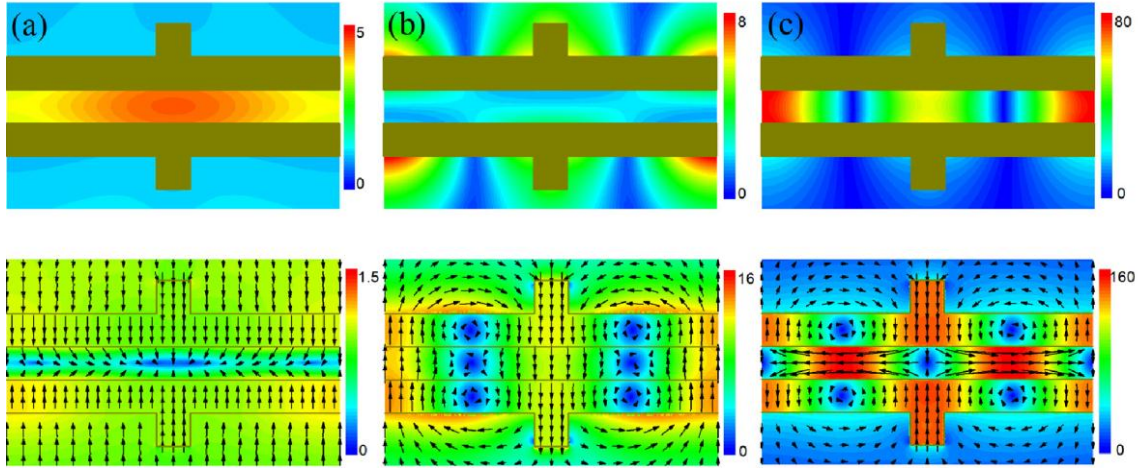

FIG. 3 (color online). Normalized near-field distributions within a single period for the three resonances $A$ (a), $B$ (b), and $C$ (c), respectively, where the colors display the amplitudes of the pressure (upper) and velocity (lower) fields, and the arrows indicate the directions of (instant) local velocities. Note that the velocity field inside the solid plates in (c) is amplified intentionally by three times to illustrate the flexural motion of the plates.

For the case of resonance $A$, the upper panel in Fig. 3(a) displays an evident pressure concentration in between the two structured plates. The pressure distribution stems from the nature of the resonance. As demonstrated by the velocity field in the lower panel (see arrows), the solid plates vibrate in the counter direction like rigid bodies, which results in a remarkable shrinking-expanding of the liquid volume and thus contributes a strong pressure field localized inside the gap. To more clearly illustrate how the (enhanced) repulsive AIMF is produced, in Fig. 4(a) we present the normalized intensities for the pressure and velocity fields distributed at $y=0$ plane. It is observed that both components of the velocity intensities are much smaller than the pressure intensity, which gives rise to the repulsive interaction as anticipated in Eq. (2). Without data presented here, similar field distributions (corresponding to the broad AIMF peak indicated by the red dashed line in Fig. 2) remain in the unstructured parallel-plate system, except for the soft pressure variation along $x$ direction due to the scattering of periodic bumps. In fact, such resonant morphologies (with nearly constant pressure and vanishing velocity) resemble that emerging in the wall-slab system [35], considering a mirror operation with respect to the rigid wall. Based on a similar reasoning, for the unstructured parallel-plate system the dimensionless AIMF can be estimated by the formula $Y_m = (\rho_s / \rho_0)(h/d) \approx 7.67$ at the normalized resonant wavelength $\lambda_0 / L = \pi L^{-1} \sqrt{2hd\rho_s / \rho_0} \approx 1.23$ [35,39], which agrees well with the numerical AIMF $Y_m = 8.07$ at $\lambda_0 / L \approx 1.27$. For the structured plates, both the AIMF ($Y_m = 8.55$) and the resonant wavelength ($\lambda_0 / L \approx 1.32$) increase slightly, because of the effective increase of the mass density by the additional square bumps.

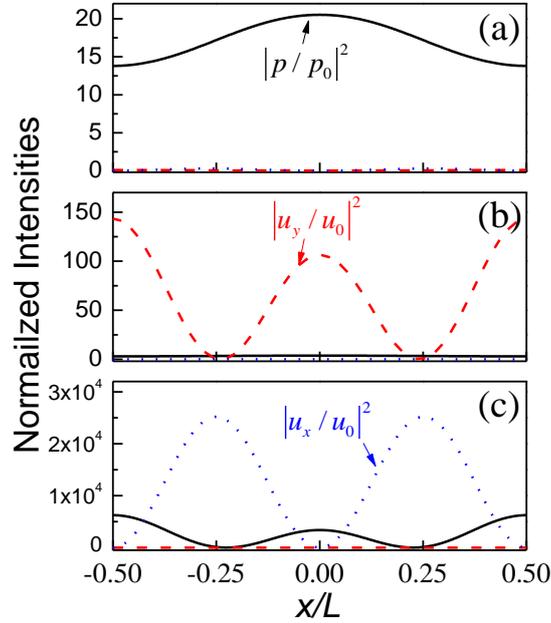

Fig. 4 (color online). The normalized intensity quantities $|p/p_0|^2$ (black solid line), $|u_x/u_0|^2$ (red dashed line), and $|u_y/u_0|^2$ (blue dotted line) distributed on the $y = 0$ plane, where (a), (b), and (c) correspond to the resonances *A*, *B*, and *C*, respectively.

The particular interest is focused on the sharply peaked AIMFs at the resonances *B* and *C*, which are unique in the structured system. As a whole, both pressure fields displayed in Figs. 3(b) and 3(c) exhibit certain symmetries with respect to the middle plane $y = 0$ if neglecting the small deviation due to the asymmetric external excitation. This feature is also exhibited in the velocity fields. Different from the resonance *A*, the velocity fields for the resonances *B* and *C* demonstrate typical flexural motions of the solid plates, which are featured by the vibration vortices pinning at stationary points. Similar resonant morphologies have been observed in the single-plate structure immersed in water, associated with extraordinary acoustic transmission phenomena [36,37]. Recently, the single-plate structure has been employed to trap particles by taking advantage of the strong acoustic field localized near the solid plate [40]. The essential physics of such vibration morphologies lies in the resonant excitation of the (nonleaky) fundamental asymmetric Lamb modes associated with flexural motions in plates.

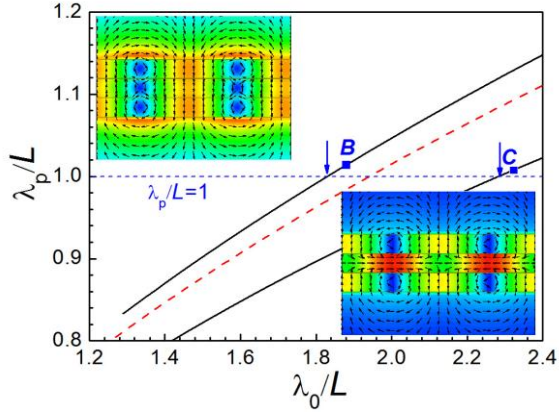

FIG. 5 (color online). Dispersion relation (black solid lines) for a pair of identical planar steel plates, comparing to that of a single steel plate (red dashed line), where $\lambda_0/L$ and $\lambda_p/L$ stand for the dimensionless wavelengths for water and plate modes, respectively. The blue squares indicate the wavelengths (in water) for the resonances *B* and *C* that occur in the structured system, slightly deviating from those predicted by the unstructured system $\lambda_p/L=1$ (see arrows). The insets display the eigen-fields for the coupled flexural modes denoted by arrows, which demonstrate obviously the in-phase and anti-phase couplings.

To further identify the resonances *B* and *C* for the double-plate structure, we calculate the dispersion relation for a pair of identical unstructured plates with thickness and distance $d=h=L/10$, together with the dispersion relation of a single planar plate for comparison. The dispersion curves are presented in Fig. 5, with the black solid lines and red dashed line for the cases of double-plate and single-plate, respectively. It is observed that the flexural mode for the single-plate now splits up due to the coupling between plates. The short wavelength branch slightly deviates from the dispersion curve of the single-plate, indicating a weak plate-plate coupling, whereas the long wavelength branch deviates more severely, indicating a stronger plate-plate coupling. This is directly linked to the vibration morphologies of the coupled flexural modes, as to be explained later. Note that these flexural modes are essentially nonleaky, i.e. $\lambda_p < \lambda_0$, with $\lambda_p$ denoting the wavelength of plate modes. They cannot be excited in the unstructured system due to the momentum mismatch with respect to the external plane wave. For the structured system, however, these modes can be resonantly excited through the additional momentum supplied by periodical bumps. The resonant wavelengths can be estimated by $\lambda_p / L = 1$ **[36]**, which is particular precise in the weak scattering limit. This estimation gives rise to resonant wavelengths in water $\lambda_0 / L = 1.83$ and $2.28$ (as denoted by arrows in Fig. 5), which are very close to the wavelengths for the resonances *B* and *C*, i.e. $\lambda_0 / L = 1.88$ and $2.31$ (as marked by squares in Fig. 5). The insets present the eigen-field distributions for the two divided flexural modes, which considerably resemble the velocity field patterns displayed in Figs. 3(b) and 3(c). Now we can conclude definitely that the resonance *B* comes from the excitation of the in-phase coupled flexural mode, whereas the resonance *C* stems from the anti-phase coupled flexural mode.

The signs of the AIMFs for the resonances *B* and *C* are closely related to the corresponding field morphologies. For the case of *B* that corresponds to the resonant excitation of the in-phase coupled flexural mode, the velocity field in Fig. 3(b) or Fig. 5 shows an odd symmetry with respect to the middle plane of the system, associated with vanishing horizontal velocity component $u_x$. Therefore, the AIMF can be reduced to $(2L)^{-1} \int_{-L/2}^{L/2} \left( |p/p_0|^2 + |u_y/u_0|^2 \right) dx$, which is always positive and explains the repulsive interaction of the resonance *B*. Different from the case of *A*, where the AIMF is enhanced through the amplified pressure field, here the AIMF enhancement is mostly determined by the drastically boosted vertical velocity component $u_y$. This can be seen more clearly in Fig. 4(b), the normalized intensities of the pressure and velocity fields distributed in the $y=0$ plane. For the case of *C* that corresponds to the resonant excitation of the anti-phase coupled flexural mode, the velocity field in Fig. 3(c) or Fig. 5 shows an even symmetry with respect to the $y=0$ plane, associated with vanishing vertical component $u_y$. This results in a simplified AIMF expression, i.e. $(2L)^{-1}\int_{-L/2}^{L/2}\left(|p/p_0|^2-|u_x/u_0|^2\right)dx$. From Fig. 4(c) it is observed that the contribution from the intensity term $|u_x/u_0|^2$ completely dominates over that from $|p/p_0|^2$, although the latter is large as well. It eventually gives rise to the huge attractive interaction between the two plates for the resonance *C*.

## V. Parameter Dependences of the Acoustic-Induced Strong Interactions

In what follows, we discuss the parameter dependences of the enhanced AIMFs for the three kinds of resonances. For the resonance *A*, the AIMF can be simply estimated as $Y_m=(\rho_s/\rho_0)(h/d)$, associated with the normalized resonant wavelength $\lambda_0/L=\pi L^{-1}\sqrt{2hd\rho_s/\rho_0}$. In these explicit expressions, there are involving only the density ratio $\rho_s/\rho_0$ and two geometric parameters $h$ and $d$. The elastic moduli (characterizing the solid's elasticity) does not appear in the formulae since the resonance *A* stems from the counter motion of the whole solid plates. For airborne sound in Ref. [35], a huge density ratio can be attained, which gives rise to a strong repulsive force at deep subwavelength. For the water background considered here, the range of the density ratio is limited by the existed materials, roughly in the order of one to ten. The further enhancement of AIMF depends on the geometry ratio $h/d$, as to be confirmed numerically below.

For the resonances *B* and *C*, which correspond to the flexural motions of the solid plates, the situation becomes much more complex since all geometric and material parameters take effects. Only qualitative predictions can be made for the resonant wavelengths: the resonances occur towards the longer wavelengths as the solid plates turn softer or turn thinner. A full-parameter exploration is tedious and beyond the scope here. With the other geometries invariant, below we discuss in details the dependences on the plate-plate distance $d$, which can be flexibly tuned for an already fabricated double-plate structure.

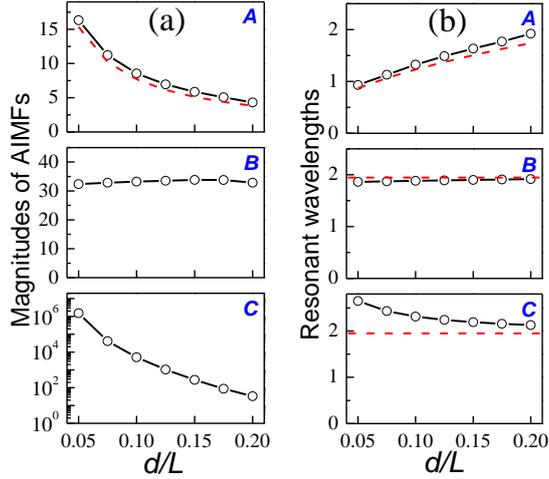

FIG. 6 (color online). (a) The magnitudes of AIMFs for the three resonances marked by $A$, $B$, and $C$, varying as the distance between the two structured plates. (b) The corresponding resonant wavelengths (normalized by $L$), where the horizontal red dashed lines present the flexural modes predicted from the single plate. For the resonance $A$, both the AIMF magnitude (a) and the wavelength (b) are compared with the analytical predictions by red dashed lines.

In Figs. 6(a) and 6(b) we present the distance dependences of the dimensionless AIMF magnitudes and wavelengths for the three resonances. It is observed that for the resonance $A$, as $d$ increases the AIMF magnitude decreases with a tendency of $\propto 1/\sqrt{d}$, accompanying with a growth of the wavelength as $\propto \sqrt{d}$. This confirms reasonably the above formulae (see red dashed lines), where the quantitative deviations partly come from the approximation used in the analytical derivation, and partly from the effective increase of the plate's mass density, as stated previously. For the resonance $B$, the magnitude of the AIMF varies slowly with $d$, associated with small variation of the resonant wavelength. The latter is very close to that predicted from the dispersion of the single-plate (see red dashed line in Fig. 6(b)). It is strikingly different for the resonance *C*: the magnitude of the AIMF grows dramatically as the reduction of $d$, associated with a resonant wavelength considerably deviated from the single-plate prediction. The difference between the cases *B* and *C* is direct connected with their resonant morphologies: the in-phase coupling (*B*) is much weaker than that of anti-phase coupling (*C*). For the former case, the in-phase flexural motions of the plates can occur almost independently, as consistent with the moderate pressure distribution manifested in Fig. 3(b). For the latter case, however, the anti-phase flexural motions may produce a rapid change of the local volume of the liquid in between plates, especially for the situation of small distance. This coincides with the strong pressure distribution in Fig. 3(c) as well, in which the fast transversal variation of pressure leads to huge horizontal component of the velocity. The distance-sensitivity of the attractive interaction could be convenient to design acoustic sensors.

## VI. Conclusions

In summary, here we present a comprehensive study on the acoustic-induced interactions between a pair of elastic plate structures. Enhanced interactions (either repulsions or attractions) have been observed in subwavelength regime. These phenomena, always linked with strong local fields, are closely related with distinct resonances. In particular, the strikingly amplified interactions that occur uniquely in the structured parallel plates are found to be originated from resonant excitation of the coupled flexural modes inherent in the elastic plates. The in-phase coupling leads to repulsive interactions which are rather stable with varied plate-plate gaps, whereas the anti-phase coupling contributes to gap-sensitive attractive interactions. Besides contactless manipulations in macroscopic scale, these findings could also be applicable in designing acoustic sensors, such as to detect weak sound signals or distance variations.

**Acknowledgments**

This work is supported by the National Natural Science Foundation of China (Grant Nos. 11374233, 11174225, 11004155, and J1210061); Program for New Century Excellent Talents (NCET-11-0398).